# THE STATUS OF MODERN FIVE-DIMENSIONAL GRAVITY

## (A Short Review: Why Physics Needs the Fifth Dimension)

Paul S. Wesson

Dept. of Physics and Astronomy, University of Waterloo, Waterloo, ON, N2L 3G1, Canada.

Abstract:  Recent criticism of higher-dimensional extensions of Einstein's theory is considered. This may have some justification as regards string theory, but is misguided as applied to five-dimensional theories with a large extra dimension.  Such theories smoothly embed general relativity, ensuring recovery of the latter's observational support.  When the embedding of spacetime is carried out in accordance with Campbell's theorem, the resulting 5D theory naturally explains the origin of classical matter and vacuum energy.  Also, constraints on the equations of motion near a high-energy surface or membrane in the 5D manifold lead to quantization and quantum uncertainty.  These are major returns on the modest investment of one extra dimension.  Instead of fruitless bickering about whether it is possible to "see" the fifth dimension, it is suggested that it be treated on par with other concepts of physics, such as time.  The main criterion for the acceptance of a fifth dimension (or not) should be its usefulness.

# THE STATUS OF MODERN FIVE-DIMENSIONAL GRAVITY

1.      Introduction and Motivation

The search for a unified account of gravity and the interactions of particles has led to an enormous log-jam of theoretical papers, while the number of clear-cut observations in favour of extra dimensions remains at zero. Not surprisingly, a distinct lobby has appeared whose aim appears to be to embarrass the proponents of extra dimensions into silence. Consider, for example, the titles of two books: *The Trouble With Physics* by Smolin [1] and *Not Even Wrong* by Woit [2]. There are also technical studies, which suggest that there are no extra dimensions to be found, no matter how small [3]. I wish, however, to prick this balloon of gloom and argue that there are tangible benefits to having one (at least) extra dimension.

The extension of Einstein's general theory of relativity from 4 to 5 dimensions, as a means of unifying gravity with electromagnetism, was suggested by Kaluza in 1921. By making the fifth dimension tiny, this approach was altered so as to include quantum effects by Klein in 1926. In more recent times, diverse choices for the number N of dimensions have been made in order to incorporate various aspects of physics. These include N=10 so as to address the vacuum fields of particles via supersymmetry, and a renewed interest in N=5 using a non-small extra dimension that is connected to the masses of particles and the origin of matter. This latter approach is known as Membrane theory or Space-Time-Matter theory, depending on its application [4-6]. The second version will be used below, but the two formulations are mathematically similar. The theory is physically very appealing: field equations for apparent vacuum in 5D can be proven to yield Einstein's field equations in 4D *with matter*, a result guaranteed by an embedding theorem of differential geometry due to Campbell [7, 8]. In this way, matter comes from geometry, a goal once espoused by Einstein. Due to the smooth embedding of general relativity, 4D cosmological models can be recovered, though with the interesting property that the embedding 5D space is flat, with no big-bang singularity. There is an especially neat form of the metric for 5D which is mathematically simple and in its pure form leads to exactly the same dynamics as Einstein's 4D theory, thereby ensuring that the 4D theory agrees with the classical tests of relativity in the solar system. This so-called canonical metric leads to new insights for the cosmological constant [9]. Departures from the pure form of the canonical metric lead to an anomalous acceleration in spacetime which is sometimes called the fifth force [10], and to a dependency of the cosmological 'constant' on the 5th. coordinate which can in principle resolve the notorious problem with the size of this parameter that plagues 4D theory [11-13]. Incidentally, the fifth force just mentioned for Space-Time-Matter theory is also present in Membrane theory. And the variation in the cosmological parameter for the one theory can steepen to form a hypersurface where the energy density of the vaccum is divergent, thereby reproducing the membrane of the second theory. These and other effects are not due to some trick of algebra, but are generic consequences of the fifth dimension.

To here, it is fair to say that our understanding of physics in four dimensions is significantly improved by the introduction of a fifth dimension. So, we can ask: Why does it not lead to some obvious new effects that can be tested? And most importantly: Why can we not *see* it?

A little thought shows that both of these questions have the same answer: To test or observe something, we have to know what it *is*. Unfortunately, while we can label the extra dimension with some convenient symbol, this does not tell us about its physical nature, or how it interacts with the other components of spacetime, or under what circumstances it can be made to stand out.

I will try, however, to illuminate these issues in the next section. The only thing I will assume at the outset is that the fifth dimension can be labelled by a coordinate and treated on the same footing as the usual four coordinates of spacetime ($x^\gamma = x^0$, $x^{123}$ for time and ordinary 3D space). That is, I will assume that the extra dimension can be labelled as a length, perhaps by the appropriate use of one or more of the constants of physics as when the speed of light is used to convert the time to a length $x^0 \equiv ct$. The latter constant, plus Newton's constant $G$ and Planck's constant $h$, will usually be absorbed by a suitable choice of units, except where they are left explicit to aid understanding. To avoid confusion with the labels for ordinary 3D space (xyz), and in order not to prejudice its physical identification, I will take the extra coordinate to be $x^4 \equiv l$. This differs from the standard usage in Membrane theory ($x^4 \equiv y$). However, other aspects of the terminology can be taken as standard, unless noted otherwise.

## 2.   The Re-interpretation of Basic Physics using the Fifth Dimension

In this section I wish to outline a working theory of the fifth dimension, drawing on results in the literature [6-13], and use it to derive some new effects. It should be recalled that a major motivation for introducing extra dimensions is unification. It will be seen that the present model not only gives the cosmological consequences mentioned in section 1, but also leads to novel effects to do with quantum mechanics.

To describe physics in an N-dimensional world needs N coordinates. And while all coordinates are allowed (covariance), the working can be heavy or light depending on how they are chosen. For laboratory physics, it is conventional to choose Cartesian coordinates ($x, y, z$). But natural systems, from an atom to a cluster of galaxies, are better described by spherical ones ($r, \theta, \phi$). As a simple model, consider the Earth and an observer who walks around the equator. Visualizing the cross-section, the fact he is confined to the surface means that he rises or falls only by small increments in the radius $dr$, as his longitude changes by a small angle $d\phi$. The actual distance he moves along the surface is given by the familiar sum of squares: $d\sigma^2 = dr^2 + r^2 d\phi^2$. This is only two-dimensional, but its simple form captures the symmetries of the physics involved.

The symmetries of five dimensions can be similarly captured, giving what on account of the simplifications it implies is called the 5D canonical metric [9]. To do this, in the previous formula the radius $r$ is replaced by a length $l$ along the extra axis; and the increment of angle $d\phi$ is replaced by the ratio $ds/L$ where $ds$ is the interval in 4D and $L$ is a length scale (here replacing the radius of the Earth in the previous example). The result, after rearranging the terms for convenience, is the 5D line element

$$dS^2 = \frac{l^2}{L^2} ds^2 + \varepsilon dl^2, \qquad ds^2 \equiv g_{\alpha\beta}(x^\gamma, l) dx^\alpha dx^\beta \qquad . \qquad (1)$$

Here $\varepsilon = \pm 1$ is an indicator of whether the extra dimension is spacelike ($\varepsilon = -1$) or timelike ($\varepsilon = +1$), both being in principle allowed. The fifth dimension embeds the 4D interval, where the latter is expressed as usual in terms of a metric tensor and coordinate increments in spacetime. However, the metric tensor may depend not only on $x^\gamma$ but also on $x^4 = l$, in which case it may be shown that *any* 5D problem can be expressed in terms of the general canonical metric (1). In fact, the dependency of the 4D metric on the fifth coordinate is what gives rise to what is conventionally known as matter [7], this being a practical application of Campbell's embedding theorem [8]. In the special case where $g_{\alpha\beta} = g_{\alpha\beta}(x^\gamma$ only) in (1), spacetime resembles a spherically-symmetric subspace embedded in 5D. In this pure-canonical case, it can be shown by evaluating Einstein's equations for (1) that there is no ordinary matter present, but there is vacuum energy measured by a cosmological constant $\Lambda = -3\varepsilon / L^2$. It is a corollary of Campbell's theorem that all vacuum solutions of general relativity, including black holes, can be embedded in the pure canonical metric.

A more general form of (1) is obtained by noting that it retains its basic form with a constant shift $l \rightarrow (l - l_0)$ along the fifth dimension. This illustrates the property of 5D relativity known as transformity, which is the change in the form of 4D physics under 5D changes of coordinates that include the extra one $x^4 = l$. The group of 5D changes $x^A \rightarrow \bar{x}^A(x^A)$ is of course broader than the group of 4D changes $x^\alpha \rightarrow \bar{x}^\alpha(x^\alpha)$, so 4D physics is malleable to some extent. In the case of a shift, the change is mathematically small but has a notable effect physically in that the effective 4D cosmological 'constant' becomes dependent on $l$ [6, 11]. Thus a shift to the pure-canonical metric gives

$$dS^2 = \frac{(l - l_0)^2}{L^2} ds^2 + \varepsilon dl^2 \; , \qquad \Lambda = -\frac{3\varepsilon}{L^2}\left(\frac{l}{l - l_0}\right)^2 \qquad . \qquad (2)$$

There are interesting implications of this for the cosmological 'constant' problem and current models of the $\Lambda$-dominated universe [12]. But there is another consequence of (2) which turns out to be of crucial importance for particle physics: there is a divergence in $\Lambda$ at $l = l_0$. That is, there is a hypersurface in 5D where the energy density of the 4D vacuum diverges. This phenomenon arises naturally in Space-Time-Matter theory, and appears to be unavoidable. Essentially the same thing is present in Membrane theory [4, 5], where though it is introduced 'by hand'. There are other, technical differences between the $\Lambda$-surface in STM theory and the singular surface of M theory, as will be seen below. It transpires that the presence of a special hypersurface is crucial for the application of 5D theory to particle physics. But before turning to quantum mechanics, a brief discussion is necessary of field equations and equations of motion.

The field equations for 5D are commonly taken to be the analog of the 4D Einstein ones for vacuum. In terms of the Ricci tensor:

$$R_{AB} = 0 \ (A, B = 0, 123, 4) \qquad . \qquad (3)$$

These apparently empty equations actually include Einstein's 4D equations, as noted elsewhere, with matter terms dependent on $\partial g_{\alpha\beta} / \partial l$ etc. (see refs. 6, 7). Many solutions to (3) are known, but it is unnecessary to go into them here, as more interest attaches to the equations of motion. These may be obtained by extremizing the interval in the standard way, via

$$\delta \left[ \int |dS| \right] = 0 \qquad . \qquad (4)$$

The result of this is a set of 5 equations of motion [9, 10]. The first 4 are a set for spacetime, which can be written as the geodesic equation for general relativity with an anomalous acceleration or force per unit mass $f^\mu$. This acts parallel to the motion along each of the axes of spacetime, and is given by

$$f^\mu = -\frac{u^\mu}{2} (\frac{\partial g_{\alpha\beta}}{\partial l} u^\alpha u^\beta) \frac{dl}{ds} \qquad . \qquad (5)$$

It is proportional to the 4-velocity ($u^\mu \equiv dx^\mu / ds$) and to the velocity through the fifth dimension as measured in terms of the 4D proper time ($dl / ds$). It is also proportional to a scalar coupling term between the 4D frame and the extra dimension as normalized using the 4-velocities. (The standard normalization condition for massive particles in spacetime is $g_{\alpha\beta} u^\alpha u^\beta = 1$.) These proportionalities mean that $f^\mu$, which is sometimes called the fifth force, is inertial in the Einstein sense. It is akin to the centrifugal force felt by a person on a roundabout in 4D. (A quick way to appreciate the existence of this extra force is to note that in any dimension of metric-based relativity the force and velocity are orthogonal as in $f^A u_A = 0$ , so in the 5D/4D case there is necessarily an extra force because $f^4 u_4 = -f^\alpha u_\alpha \neq 0$.) While (4) and (5) above can be used to obtain the 5D equations of motion in the general case, a simpler procedure is available that gives the velocity in the extra dimension for metrics which have a simple form, like (1) or (2). Then the paths of particles may be obtained directly from the null condition $dS^2 = 0$ [6, 13]. This agrees with the null form of (3), and has been used in both STM and M theory. It means that all particles behave like photons in 5D whether or not they have rest mass in 4D. In other words, $dS^2 = 0$ replaces $ds^2 \geq 0$ as the basis for causality.

The introduction of a fifth dimension leads to new interpretations of several aspects of quantum mechanics and the resolution of some long-standing problems. As a prime example, it has never been clear why particles and waves provide alternate descriptions of small-scale phenomena. This puzzle can in principle be solved, because 5D dynamics has two modes, depending on whether the extra dimension is spacelike or timelike. The background is assumed to be vacuum, so the metric may be taken to be the shifted form of the pure-canonical one given by (2). As mentioned above, setting $dS^2 = 0$ will give the path $l = l(s)$ through the fifth dimension as a function of the 4D proper time. Depending on the nature of the extra dimension, this path is either monotonic ($\varepsilon = -1$ , spacelike) or oscillatory ($\varepsilon = +1$ , timelike). To streamline the algebra, the symbol ($i$) can be used, meaning that this factor is present for oscillatory motion but absent for monotonic motion. Thus there is a constant of the (5D) motion given by

$$\frac{1}{(l-l_0)}\frac{dl}{ds} = \frac{\pm(i)}{L} \qquad . \tag{6.1}$$

The sign choice here reflects the reversibility of the motion along the $x^4 = l$ axis. Integrating and introducing an arbitrary constant of integration $l_*$ gives the bimodal equation of motion:

$$l = l_0 + l_* \exp[\pm(i)s/L] \qquad . \tag{6.2}$$

For the monotonic mode, a particle either approaches or recedes from $l = l_0$ at a rate governed by $L$. For the oscillatory mode, a wave undergoes simple harmonic motion about $l = l_0$ with amplitude $l_*$ and wavelength $L$. This wave is in effect locked to the locus $l = l_0$. The same result as (6.2) is obtained from the more general approach based on (4), which shows that in application to particles the equation of motion for $l(s)$ is actually the Klein-Gordon wave equation, where $L$ is the Compton wavelength $h/mc$ [13]. More work is needed on the relationship between particles and waves, but it is clear that they can be regarded as the alternate modes of the motion in a vacuum-dominated 5D metric. If the modes occur together, their associated spacetime curvatures will cancel (see below). So the problem of wave-particle duality can in principle be resolved in a spacetime that is close to flat.

This conclusion is based on the approach of Space-Time-Matter theory rather than the alternative Membrane theory, and it instructive to make some comments here about the nature of the membrane in these approaches before moving on to other topics.

As mentioned in Section 1, the STM and M theories are mathematically similar. But their philosophies differ somewhat, and this is most obvious when their membranes are compared. The membrane in M theory was postulated in order to explain the relative strengths of particle interactions and gravity, the former being confined to a singular 4D hypersurface while the latter can propagate freely in 5D. This also has implications for the masses of particles (an account of both theories is given in ref. 6). In M theory, the membrane is symmetric. The membrane in STM theory is not postulated, but arises naturally from the field equations as a consequence of the shift ($l_0$) along the $x^4$ axis of the canonical metric (2). This membrane is akin to an horizon in general relativity like that familiar from a black hole. It is not singular, because the wave described by (6.2) can cross the surface $l = l_0$ (its range is from $l_0 - l_*$ to $l_0 + l_*$). Also, the membrane is not symmetric, as can be seen from (2) for the energy density of the vacuum as measured by the cosmological 'constant' $\Lambda$. Since $\Lambda = \Lambda(l)$, a spectrum of particle masses is expected, depending on the local value of the energy density of the vacuum. ( For $l \rightarrow \infty$, $|\Lambda| \rightarrow 3/L^2$ where $L$ is the scale of the 4D potential and also defines via the Compton wavelength a limiting value for particle mass.) In STM theory, modes are allowed with $\Lambda > 0$ and $\Lambda < 0$, which correspond in the simplest case to de Sitter and anti-de Sitter spacetimes embedded in a 5D manifold with spacelike and timelike extra dimensions. In M theory, the focus is on the anti-de Sitter mode only. This allows for velocities in ordinary 3D space which are superluminal, as implied by certain aspects of quantum mechanics. But a more complete account of 5D dynamics is provided by the null-geodesic condition $dS^2 = 0$ of STM theory.

It was noted above that the simplest way to account for wave-particle duality is to assume that the timelike and spacelike modes of STM theory occur together. If the theory is restricted to one extra dimension, it is possible to generalize the canonical metric (2) by introducing a scalar field ($g_{44}$) which is complex, its two modes being associated with waves and particles [6]. Another possibility is to extend STM theory to more than five dimensions, something which has already been considered for M theory. In fact the STM and M approaches may turn out to be complimentary if the manifold is extended to $N > 5D$. However, while this may be feasible algebraically, it is difficult to justify physically, unless a practical meaning can be given to the higher dimensions. In this regard, 5D is the basic extension of 4D general relativity, and in the STM approach the extra dimension is a representation of matter and energy.

The material I have discussed to here has been known to workers in the field for several years. But I now wish to consider two subjects, namely quantization and uncertainty, which find clarification only now.

Quantization is the dominant feature of matter at short distances, and it is reasonable to expect that the canonical 5D metric will explain at least the main features of quantum physics. There are several ways to construct models of particles in 5D, some of which are quite complicated [6, 13]. But it seems to have eluded notice that quantization is simple and inevitable as a particle/wave approaches the special surface $l = l_0$ in the canonical metric (2). To see this, it is sufficient to take (6.1) and rewrite it, dropping non-essential factors, to give

$$\frac{dl}{(l - l_0)} = \frac{ds}{L} = \frac{mc\,ds}{h} \quad . \tag{7}$$

Here the identification for $L = h/mc$ noted above has been used, where $m$ is the mass of the particle associated with the wave of (6.2). As the wave approaches $l_0$, both $dl$ and $(l - l_0)$ approach zero with the ratio $dl/(l - l_0) \to 1$. Then by (7) $mc\,ds \to h$, giving the usual quantum rule. This is a kind of automatic quantization, made obligatory by the mechanics. To obtain a multiple $n$ of quantum steps it is merely necessary to change $L$ to $nL$ in the preceding analysis. (In the more general case alluded to above where $g_{44}$ is a complex scalar field, the same result follows if the eigenstates of the field are given by $g_{44} = n^2$.) The inference is that quantization with the old wave mechanics rule $mc\,ds = nh$ is due to being near a special hypersurface in the fifth dimension.

The quantum uncertainty relation, it turns out, also has to do with near $-l_0$ behaviour. In that region, the anomalous force per unit mass (5) due to the extra dimension can become large (assuming the coordinates are not chosen especially to remove it). It can be evaluated for the canonical-type metric (2) by using $g_{\alpha\beta} = (l - l_0)^2 L^{-2} \tilde{g}_{\alpha\beta}(x^\gamma)$ where $\tilde{g}_{\alpha\beta}(x^\gamma)$ is the metric tensor of spacetime. Then $\partial g_{\alpha\beta} / \partial l = 2(l - l_0)^{-1} g_{\alpha\beta}$. Assuming that the 4-velocities are normalized as noted before via $g_{\alpha\beta} u^\alpha u^\beta = 1$ the scalar coupling term in (5) is $(\partial g_{\alpha\beta} / \partial l) u^\alpha u^\beta = 2(l - l_0)^{-1}$. And (5) gives

$$f^\mu = -\frac{u^\mu}{(l - l_0)} \frac{dl}{ds} = \mp \frac{(i)}{L} u^\mu \qquad , \tag{8}$$

where (6.1) has been used. The sign choice here has to do with reversibility, and the $(i)$ arises because the acceleration $f^\mu = du^\mu / ds$ and the 4-velocity $u^\mu$ are necessarily out of phase for simple harmonic motion (see above). Neither thing is important for the analysis, so they may be dropped. Then (8) reads

$$\frac{du^\mu}{ds} = \frac{1}{L} \frac{dx^\mu}{ds} \quad \text{or} \quad du^\mu = \frac{dx^\mu}{L} \qquad . \tag{9}$$

This last equation can be employed to form the scalar quantity $du^\mu dx_\mu = ds^2 / L$. This may in turn be re-expressed using (7) and the condition $dl / (l - l_0) \to 1$ wherein $ds / L \to 1$. The result is $du^\mu dx_\mu = L$. Substituting for $L = h / mc$ from before, and replacing the change in velocity by the change in momentum, gives

$$dp^u dx_\mu = h \qquad .$$  (10)

This resembles the usual quantum uncertainty relation. Strictly speaking, however, (10) refers only to the mismatch in momentum and position due to the fifth force (5) arising from the extra dimension. Extra forces, of a 4D nature, will modify the offset.

3.      Discussion and Conclusion

Einstein's opinion about extra dimensions oscillated through his lifetime. At one stage, writing with Bergmann in 1938, he stated: "We ascribe physical reality to the fifth dimension" [14]. It is remarkable, for both physics and psychology, that after so many years there is still no consensus (let alone empirical proof) about the fifth dimension. Before delving into the reasons, it is instructive to review what has been shown above about the achievements of one well-studied approach to the problem.

There are many ways to construct a map with spacetime and an extra dimension. An obvious choice is to use Cartesian coordinates, where the total element or distance is a sum of the squares of the increments in spacetime and an extra orthogonal dimension, like a right-angled triangle. But an option preferred by nature is to imagine that spacetime is measured around the perimeter of a circle, with the fifth dimension being like the radius. It is important to realize that in this map, the actual shape of spacetime is not necessarily a circle, since appropriately-placed points in circular polar coordinates can describe any shape. However, it *is* true that the 'default' shape is a circle. This is the case for the so-called 5D canonical metric (1), where spacetime is embedded as a kind of spherical surface, like the surface of the Earth in ordinary 3D space. The physical justification for this metric is that it embeds all vacuum solutions of Einstein's 4D theory, where the energy density of the vacuum is measured by the cosmological constant $\Lambda$ . The form of the canonical metric is preserved when there is added a shift along the extra dimension, and surprisingly the new $\Lambda$ is not constant and indeed diverges at a hypersurface defined by a special value of the extra coordinate as in (2). This surface enters naturally in Space-Time-Matter theory, but is put in artificially at the foundation of Membrane theory. Irrespective of the version of 5D relativity, it may as well be called a membrane. The physics everywhere in 5D space – far from and near to the membrane – is given by the field equations (3). These include Einstein's field equations of 4D general relativity, together with matter. This originates because the 4D potentials now have an extra degree of freedom associated with the fifth dimension, and the 'force' associated with the gradient in that direction manifests itself as energy (matter). This and other aspects of the embedding of 4D in 5D are governed by Campbell's theorem. The paths of individual particles through the 5D manifold can be obtained by carrying out the variational exercise summed up in (4). This shows that, in general, an extra force per unit mass (5) acts in spacetime. A further consequence of the extension of the manifold is that the conventional definition of causality using the 4D proper time ( $ds^2 \geq 0$ ) can be replaced by a null condition on the 5D interval ( $dS^2 = 0$ ). So in 5D all particles behave like photons and everything in the universe is in causal contact with everything else.

On the small as opposed to large scale, the vacuum and the membrane typical of it play a dominant role. The equation of motion in the extra dimension shows a monotonic mode and an oscillatory mode, dependent on the sign of the extra metric coefficient, as in (6.2). It is natural to identify these with the paths of particles and waves, and the manifold remains 4D flat if both modes are realized. (The details of this are still obscure, though it could be that the extra metric coefficient is not just unity as in the canonical metric but represents a complex scalar field with the membrane a breakdown in it.) Quantization is a characteristic of the approach to the membrane, whose properties are such that the motion is constrained as in (7), resulting in the standard

Planck-type rule. The quantum uncertainty relation is also connected with the influence of the membrane, because the motion in the extra dimension is linked to it. The analysis of (8)-(10) shows that the momentum and the position became autocorrelated in the manner of the standard Heisenberg relation. It should be said that while quantization and uncertainty can in principle be explained by a 5D manifold with a special hypersurface, other aspects of the membrane remain to be clarified.

Even so, the list of achievements due to the introduction of one extra dimension is impressive. So why is the fifth dimension not more widely accepted, and why has it recently come under attack? The criticism is strongest against theories with many dimensions (i.e. N>5), particularly string theory, but does include modern versions of the old 5D Kaluza-Klein theory [1-3]. Perhaps the most telling objection is that after many years of investigation there is still no empirical proof of the existence of even one extra dimension. This is a valid position, especially as regards approaches like string theory which have a significantly different basis than general relativity. But the lack of experimental support for extra dimensions is not a strong enough argument to justify the abandonment of modern theories which use only one extra dimension and smoothly embed general relativity by virtue of Campbell's theorem. For such approaches, it is not a question of right or wrong, but rather one of degree of support. Some workers would argue that explaining the origin of matter is enough of a return to be worth the investment of belief in one extra dimension. Nevertheless, the comment is commonly heard that it is difficult to believe in a fifth dimension when it cannot be "seen". This objection, on closer consideration, should be regarded as suspect. Newton could not "see" gravity when he formulated the inverse-square law for it. And Einstein could not "see" the curvature of spacetime when he used it to re-express the gravitational force.  In physics, the question of whether something can be seen or not is frequently irrelevant.

It was pointed out by Eddington, in the early days of general relativity and quantum mechanics, that whether something can be seen is a hopelessly subjective way of deciding if it belongs in physics.  He went on to develop a comprehensive philosophy of science [15].  This has been re-examined in relation to modern physics, with the conclusion that the subject progresses not so much by seeing new things as by improving its logical framework [16].  While that view may not be universally accepted, it is shown by the propensity in physics for constructing models.  This mode of progress is taking over cosmology, and has long been the standard in quantum theory.  It has been shown above that one extra dimension – of the appropriate sort – can provide improved understanding on both the large and small scales.  This applies to the nature of the vacuum and the origin of classical matter, as well as the reasons for quantization and quantum uncertainty.  It can be argued that the introduction of the fifth dimension has significantly improved the logical consistency of physics.  And further improvement may be expected and would certainly be welcome.  At present, quantum theory in particular is in a disordered state, its formalism littered with the paradoxes of old models: wave-particle duality, pilot waves, decoherence, collapse of the wave function, loss of information, time-ordering, zero-point fields, renormalization… [17-20]. These and other problems, it is reasonable to expect, should be cleared up by any new approach.

I think it is fair to suggest that if the fifth dimension succeeds in rationalizing our understanding of quantum mechanics, then it should be provisionally accepted as in some sense existing.  This over the objections of certain workers who may wish to continue arguing that they cannot believe in something unless they can "see" it.  After all, we cannot "see" time, but it en-ters almost every equation of physics. I suggest that the fifth dimension be treated – like time – as a concept, whose acceptance (or not) should be decided by how useful it is.

Acknowledgements

Although their views may not coincide exactly with mine, I am grateful for discussions with several members of the Space-Time-Matter group (5Dstm.org).